\documentclass[reprint,prl,superscriptaddress]{revtex4-1}
\usepackage{bm}
\usepackage{graphicx}
\usepackage[large]{subfigure}
\usepackage{amssymb, amsmath}
\usepackage[amssymb]{SIunits}
\usepackage{tabularx}
\usepackage{color}
\renewcommand\section[1]{\emph{#1}---}
\renewcommand\subsection[1]{\emph{#1}---}

\begin{document}

\title{Cosmological Parameters from Pre-Planck CMB Measurements}
\author{Erminia~Calabrese}\affiliation{Sub-department of Astrophysics, University of Oxford, Keble Road, Oxford OX1 3RH, UK}
\author{Ren\'ee~A.~Hlozek}\affiliation{Dept. of Astrophysical Sciences, Peyton Hall, Princeton University, Princeton, NJ 08544, USA}
\author{Nick~Battaglia}\affiliation{Department of Physics, Carnegie Mellon University, Pittsburgh, PA 15213, USA}
\author{Elia~S.~Battistelli}\affiliation{Department of Physics, University of Rome `Sapienza', Piazzale Aldo Moro 5, I-00185 Rome, Italy}
\author{J.~Richard~Bond}\affiliation{CITA, University of Toronto, Toronto, ON M5S 3H8, Canada}
\author{Jens~Chluba}\affiliation{Johns Hopkins University, 3400 N. Charles St., Baltimore, MD 21218-2686, USA}
\author{Devin~Crichton}\affiliation{Johns Hopkins University, 3400 N. Charles St., Baltimore, MD 21218-2686, USA}
\author{Sudeep~Das}\affiliation{High Energy Physics Division, Argonne National Laboratory, 9700 S Cass Avenue, Lemont, IL 60439, USA}\affiliation{BCCP, LBL and Department of Physics, University of California, Berkeley, CA 94720, USA}
\author{Mark~J.~Devlin}\affiliation{Department of Physics and Astronomy, University of Pennsylvania, 209 South 33rd St., Philadelphia,PA 19104,USA}
\author{Joanna~Dunkley}\affiliation{Sub-department of Astrophysics, University of Oxford, Keble Road, Oxford OX1 3RH, UK}
\author{Rolando~D\"{u}nner}\affiliation{Departamento de Astronom{\'{i}}a y Astrof{\'{i}}sica, Pontific\'{i}a Universidad Cat\'{o}lica de Chile, Casilla 306, Santiago 22, Chile}
\author{Marzieh~Farhang}\affiliation{CITA, University of Toronto, Toronto, ON M5S 3H8, Canada}\affiliation{Department of Astronomy and Astrophysics, University of Toronto, 50 St George , Toronto, ON, M5S 3H4}
\author{Megan~B.~Gralla}\affiliation{Johns Hopkins University, 3400 N. Charles St., Baltimore, MD 21218-2686, USA}
\author{Amir~Hajian}\affiliation{CITA, University of Toronto, Toronto, ON M5S 3H8, Canada} 
\author{Mark~Halpern}\affiliation{Department of Physics and Astronomy, University of British Columbia, Vancouver, BC V6T 1Z4, Canada}
\author{Matthew~Hasselfield}\affiliation{Dept. of Astrophysical Sciences, Peyton Hall, Princeton University, Princeton, NJ 08544, USA}\affiliation{Department of Physics and Astronomy, University of British Columbia, Vancouver, BC V6T 1Z4, Canada}
\author{Adam~D.~Hincks}\affiliation{CITA, University of Toronto, Toronto, ON M5S 3H8, Canada} 
\author{Kent~D.~Irwin}\affiliation{NIST Quantum Devices Group, 325 Broadway Mailcode 817.03, Boulder, CO 80305, USA}
\author{Arthur~Kosowsky}\affiliation{Department of Physics and Astronomy, University of Pittsburgh, Pittsburgh, PA 15260, USA}
\author{Thibaut~Louis}\affiliation{Sub-department of Astrophysics, University of Oxford, Keble Road, Oxford OX1 3RH, UK}
\author{Tobias~A.~Marriage}\affiliation{Johns Hopkins University, 3400 N. Charles St., Baltimore, MD 21218-2686, USA}\affiliation{Dept. of Astrophysical Sciences, Peyton Hall, Princeton University, Princeton, NJ 08544, USA}\affiliation{Joseph Henry Laboratories of Physics, Jadwin Hall, Princeton University, Princeton, NJ 08544,USA}  
\author{Kavilan~Moodley}\affiliation{Astrophysics and Cosmology Research Unit, School of
Mathematical Sciences, University of KwaZulu-Natal, Durban, 4041, South Africa}
\author{Laura~Newburgh}\affiliation{Joseph Henry Laboratories of Physics, Jadwin Hall, Princeton University, Princeton, NJ 08544,USA}
\author{Michael~D.~Niemack}\affiliation{Joseph Henry Laboratories of Physics, Jadwin Hall, Princeton University, Princeton, NJ 08544,USA}\affiliation{NIST Quantum Devices Group, 325 Broadway Mailcode 817.03, Boulder, CO 80305, USA}\affiliation{Department of Physics, Cornell University, Ithaca, NY, USA 14853}
\author{Michael~R.~Nolta}\affiliation{CITA, University of Toronto, Toronto, ON M5S 3H8, Canada}
\author{Lyman~A.~Page}\affiliation{Joseph Henry Laboratories of Physics, Jadwin Hall, Princeton University, Princeton, NJ 08544,USA}  
\author{Neelima~Sehgal}\affiliation{Physics and Astronomy Department, Stony Brook University, Stony Brook, NY 11794-3800, USA}
\author{Blake~D.~Sherwin}\affiliation{Joseph Henry Laboratories of Physics, Jadwin Hall, Princeton University, Princeton, NJ 08544,USA}  
\author{Jonathan~L.~Sievers}\affiliation{Joseph Henry Laboratories of Physics, Jadwin Hall, Princeton University, Princeton, NJ 08544,USA}  
\author{Crist\'obal~Sif\'on}\affiliation{Leiden Observatory, Leiden University, PO Box 9513, NL-2300 RA Leiden, Netherlands}
\author{David~N.~Spergel}\affiliation{Dept. of Astrophysical Sciences, Peyton Hall, Princeton University, Princeton, NJ 08544, USA}
\author{Suzanne~T.~Staggs}\affiliation{Joseph Henry Laboratories of Physics, Jadwin Hall, Princeton University, Princeton, NJ 08544,USA}
\author{Eric~R.~Switzer}\affiliation{CITA, University of Toronto, Toronto, ON M5S 3H8, Canada}
\author{Edward~J.~Wollack}\affiliation{NASA/Goddard Space Flight Center, Greenbelt, MD 20771, USA}

\begin{abstract}
Recent data from the WMAP, ACT and SPT experiments provide precise measurements of the cosmic microwave background temperature power spectrum over a wide range of angular scales. The combination of these observations is well fit by the standard, spatially flat $\Lambda$CDM cosmological model, constraining six free parameters to within a few percent. The scalar spectral index, $n_s = 0.9690 \pm 0.0089$, is less than unity at the 3.5$\sigma$ level, consistent with simple models of inflation. The damping tail of the power spectrum at high resolution, combined with the amplitude of gravitational lensing measured by ACT and SPT, constrains the effective number of relativistic species to be $N_{\rm{eff}}=3.28 \pm 0.40$, in agreement with the standard model's three species of light neutrinos.  
\end{abstract}
\maketitle

\section{Introduction}
It has long been appreciated that the Cosmic Microwave Background (CMB) power spectrum contains enough information to precisely determine the standard model of cosmology \cite{spergel94,jungman96,zaldarriaga97,bond97}. This promise has been realized through a series
of increasingly sensitive experiments, most recently with the WMAP satellite's 9-year full-sky observations \citep{Bennett:2012fp,Hinshaw:2012fq} and the arcminute-resolution maps from the Atacama Cosmology Telescope (ACT) \citep{Das:2013zf,Dunkley:2013vu,Sievers:2013wk} and the South Pole Telescope (SPT) \citep{Story:2012wx,Hou:2012xq}. The combination of these measurements probes the temperature power spectrum on angular scales ranging from 90 degrees to 4 arcminutes, scales at which the primary cosmological temperature fluctuations dominate. The primordial fluctuations are well approximated as a Gaussian random field \citep{Bennett:2012fp,Komatsu:2003fd}, but ACT and SPT have also detected the non-Gaussian features due to gravitational lensing of the microwave radiation by the intervening large-scale structures \citep{das/etal:2011,vanEngelen/etal:2012}. 
In this letter we present a joint analysis of the ACT, SPT, and final WMAP 9-year power spectra to obtain an estimate of the cosmological parameters from microwave background data alone. 

\section{Data and Analysis Method}\label{sec:data}
\begin{figure*}
\includegraphics[scale=0.58]{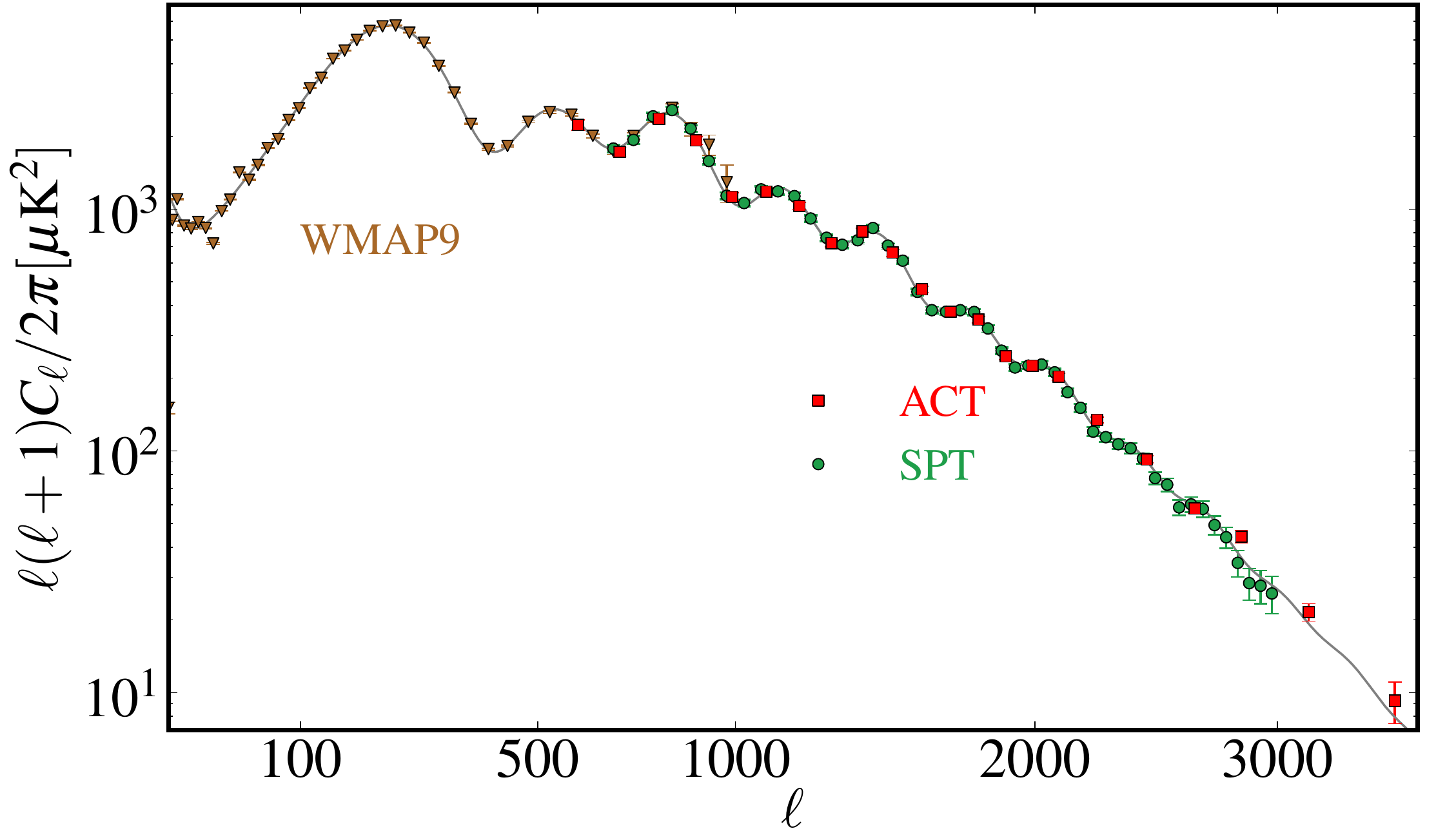}
\caption{WMAP9 temperature data and ACT and SPT CMB lensed bandpowers marginalized over secondary emissions. The ACT bandpowers are estimated separately for ACT-S and ACT-E and coadded here with an inverse variance weighting. The SPT bins are highly correlated, ($50-65\%$) at small scales, $\ell \gtrsim 2000$, due to foreground uncertainty. The correlation is about $5\%$ between neighbouring ACT bins. The solid line shows the lensed CMB best fit obtained combining the three datasets. The ACT and SPT bandpowers are available on LAMBDA (http://lambda.gsfc.nasa.gov/).
\label{fig:data}
}
\end{figure*}
In Figure \ref{fig:data} we show the compilation of CMB temperature power spectra used in this analysis. At large angular scales we use the temperature and polarization data, and associated likelihood, from the 9-year WMAP analysis \citep[hereafter WMAP9,][]{Hinshaw:2012fq}. This measures the Sachs-Wolfe plateau and the first three acoustic peaks, $2<\ell \lesssim1000$. At smaller scales, $500<\ell<3500$, we use data from ACT and SPT.

Here we follow the method introduced in \cite{Dunkley:2013vu} to estimate the primary CMB bandpowers from both sets of spectra, marginalizing over the possible additional power from Galactic and extragalactic emission, and the Sunyaev-Zel'dovich effects. 
We use a Gibbs sampling method to simultaneously estimate CMB bandpowers and a set of ten secondary parameters. For ACT we extract primary CMB bandpowers from the 148 and 218~GHz auto and cross power spectra from two regions (ACT-E and ACT-S, \cite{Das:2013zf}) of the sky \cite{binning}, taking the multi-frequency bandpowers in the range 500$<\ell<$10000. We include SPT 150~GHz data \cite{Story:2012wx} from 650$<\ell<$3000, and marginalize over a common model for secondary components \cite{secondary}. We impose a Gaussian prior of $12.3 \pm 3.5 \ \mu\rm{K}^2$ at $\ell=3000$ on the SPT radio source Poisson power, having subtracted $7\ \mu\rm{K}^2$ of cosmic infrared background Poisson power, treated separately in our likelihood, from the total expected Poisson level \citep{Story:2012wx}, \cite{w7s12}. The resulting ACT and SPT lensed bandpowers are shown in Figure \ref{fig:data}, and the secondary parameters are consistent with those reported in \cite{Dunkley:2013vu,Sievers:2013wk}. The errors shown are the diagonal elements of the covariance matrix, with the SPT calibration error removed for consistency with ACT. The full covariance matrix includes correlations due to foreground uncertainty, beam error, and the overall calibration for SPT. 

We then construct an ACT+SPT likelihood from these CMB bandpowers, which can also be used for each experiment on its own. This is a Gaussian distribution using 42 data points from ACT (21 each from ACT-E and ACT-S) and 47 from SPT, with an associated covariance matrix. For ACT we only use $\ell<$3500 bandpowers in the likelihood, where their distributions are Gaussian. When combining ACT with SPT, we use only ACT-E data to eliminate the covariance between ACT-S and SPT, which observe overlapping sky regions. 
We combine this likelihood with WMAP9, using the CosmoMC code \citep{Lewis:2002ah} to estimate cosmological parameters. 
\begin{table}
\centering
\caption{Standard $\Lambda$CDM parameters from the combination of WMAP9, ACT and SPT. \label{tbl1}}
\centering
\begin{tabular}{lccc}
\hline\hline
 Parameter & WMAP9 & WMAP9 & WMAP9\\
 & +ACT & +SPT & +ACT+SPT$^a$\\
\hline
$100\Omega_b h^2$ & $2.260 \pm 0.041$ & $2.231 \pm 0.034$ & $2.252 \pm 0.033$\\
$100\Omega_c h^2$ & $11.46 \pm 0.43$ &  $11.16 \pm 0.36$ & $11.22 \pm 0.36$\\
$100 \theta_A$ & $1.0396 \pm 0.0019$ &  $1.0422 \pm 0.0010$ & $1.0424 \pm 0.0010$\\
$\tau$ & $0.090 \pm 0.014$ & $0.082 \pm 0.013$ & $0.085 \pm 0.013$\\
$n_s$  & $0.973 \pm 0.011$ &  $0.9650 \pm 0.0093$ & $0.9690 \pm 0.0089$\\
$10^{9}\Delta_{\cal R}^2$ & $2.22 \pm 0.10$ & $2.15 \pm 0.10$ & $ 2.17 \pm 0.10$\\
\hline
$\Omega_\Lambda$$^b$  & $0.716 \pm 0.024$ & $0.737 \pm 0.019$ & $0.735 \pm 0.019$ \\
$\sigma_8$& $ 0.830\pm 0.021$ & $ 0.808\pm 0.018$ & $ 0.814\pm 0.018$ \\
$t_0$ & $13.752 \pm 0.096$ & $13.686 \pm 0.065$ & $13.665 \pm 0.063$\\
$H_0 $ & $69.7 \pm 2.0$ & $71.5 \pm 1.7 $ & $71.4 \pm 1.6$ \\
$100r_s/{D_V}_{0.57}$ &$7.50 \pm 0.17$ &$7.65 \pm 0.14$ &$7.66 \pm 0.14$  \\
$100r_s/{D_V}_{0.35}$ &$11.29 \pm 0.31$ & $11.56 \pm 0.26$ & $11.57 \pm 0.26$ \\
\hline
best fit $\chi^2$ & $7596.0$ & $7617.1$ & $7640.7$\\
\hline\hline
\footnotetext[1]{The combination ACT+SPT uses ACT-E data only.\\
\noindent
We report errors at $68\%$ confidence levels.}
\footnotetext[2]{Derived parameters: Dark energy density, the amplitude of matter fluctuations on 8 $h^{-1} \rm{Mpc}$ scales, the age of the Universe in \rm{Gyr}, the Hubble constant in units of \rm{km/s/Mpc}, and the galaxy correlation scales at redshifts 0.57 and 0.35.}
\end{tabular}
\end{table}

We consider the basic spatially-flat $\Lambda$CDM cosmological model defined by six parameters: the baryon and cold dark matter densities, $\Omega_b h^2$ and $\Omega_c h^2$; the angular scale of the acoustic horizon at decoupling, $\theta_A$; the reionization optical depth, $\tau$; the amplitude and the scalar spectral index of primordial adiabatic density perturbations, $\Delta_{\cal R}^2$  and $n_{\rm{s}}$ (at a pivot scale $k_0 = 0.05~\rm{Mpc}^{-1}$). We also extend the standard model to include a seventh parameter $N_{\rm eff}$, the effective number of relativistic species at decoupling.  The high-$\ell$ damping tail measured by ACT and SPT is particularly sensitive to this parameter.

\section{Results and discussion} \label{sec:res}
\begin{figure}
\center
\includegraphics[width=\columnwidth]{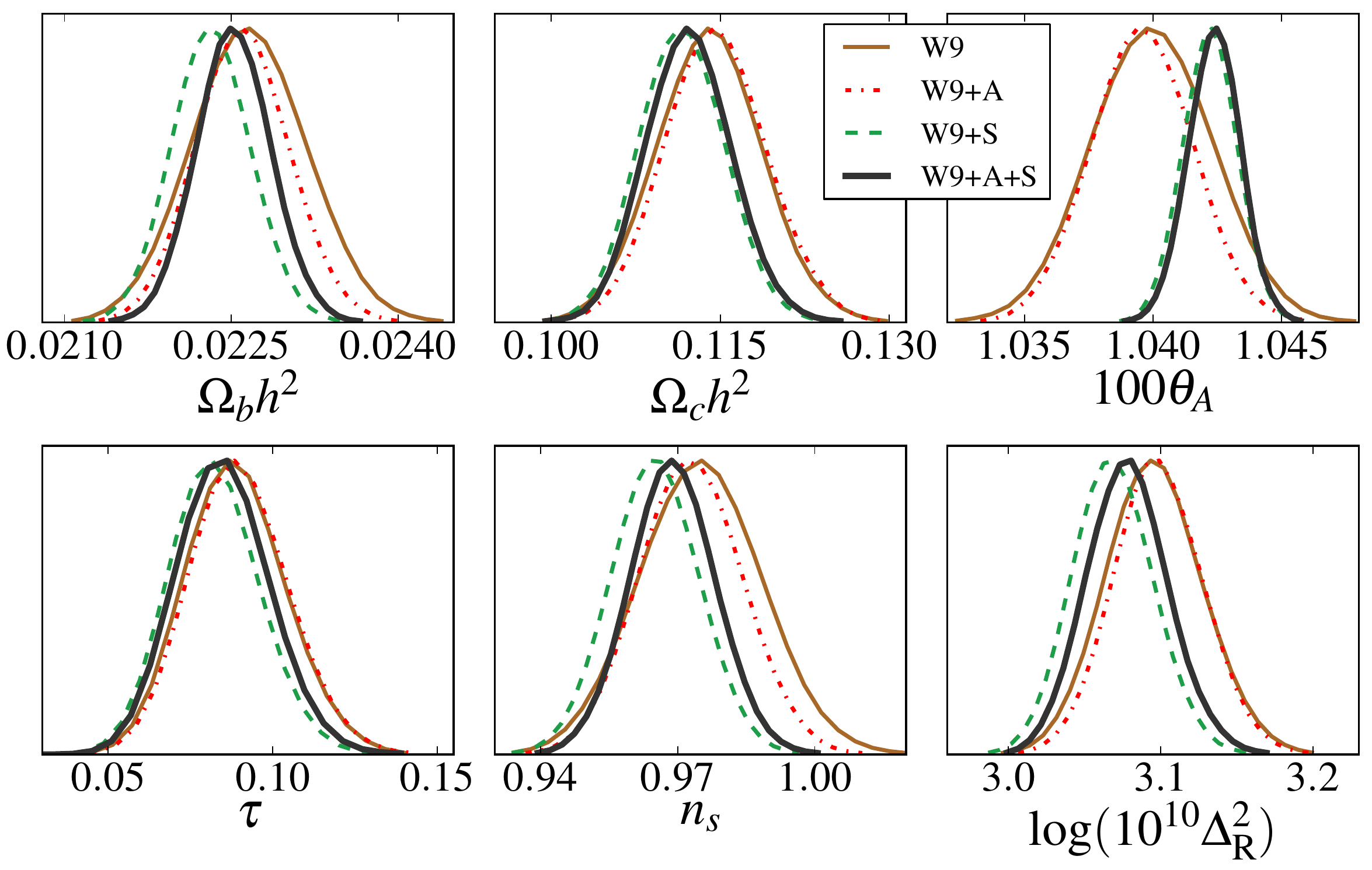}
\caption{Marginalized one-dimensional distributions for the six basic $\Lambda$CDM parameters, for combinations of WMAP9 (W9), ACT (A) and SPT (S) data.
\label{fig:lcdm}
}
\end{figure}

\begin{figure}
\center
\includegraphics[width=\columnwidth]{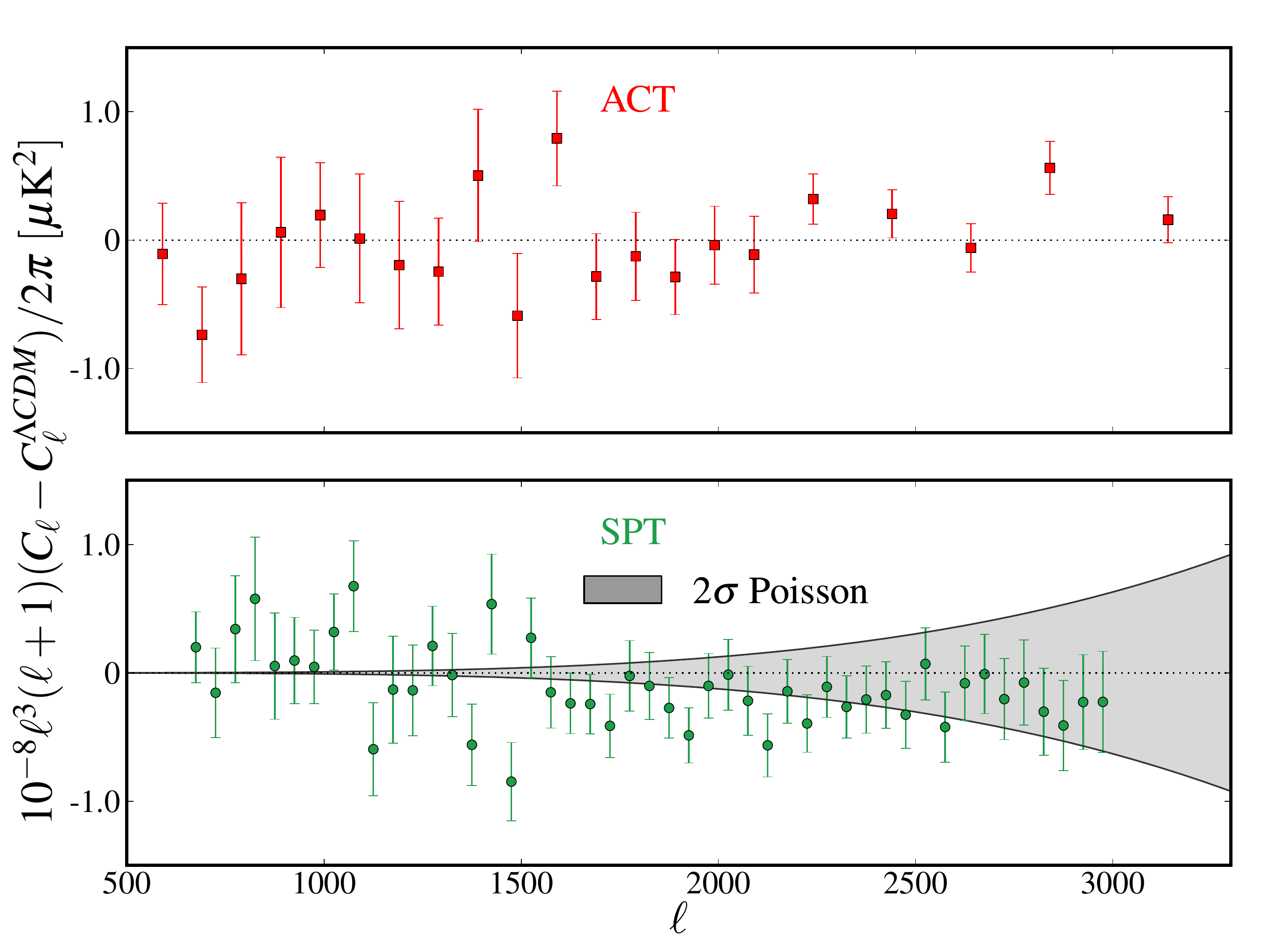}
\caption{Residual power after subtracting the same best-fitting lensed CMB model. The reduced $\chi^2$/dof for ACT is $40.1/42$ (PTE=0.55) and for SPT $55.7/47$ (PTE=0.18). We show ACT-E and ACT-S coadded residuals. The grey band in the bottom panel shows the $2\sigma$ uncertainty in the Poisson source component. Overall calibration errors are not included.
\label{fig:res}
}
\end{figure}

The simple $\Lambda$CDM model fits all the data well, with the estimated parameters shown in Table \ref{tbl1} and Figure \ref{fig:lcdm}.
We find $\chi^2$/dof$ = 37.9/42$ (Probability to Exceed, PTE = 0.65) for ACT and $53.2/47$ (PTE = 0.25) for SPT when combined individually with WMAP9, assuming the degrees of freedom equal the number of additional data points. The best-fitting parameters for ACT are all within about $1\sigma$ of the corresponding best-fitting parameters for SPT. For the combined analysis, the ACT+SPT best fit $\chi^2$/dof is $78.9/68$ (PTE $= 0.17$). Compared to the joint best-fit model, the SPT-only best-fit has $\Delta\chi^2=2.5$ worse and the ACT-only best fit (for ACT-S+ACT-E) has $\Delta\chi^2=2.2$, indicating that the common model fits both datasets. Figure \ref{fig:res} shows the residual power for the high-$\ell$ datasets after subtracting the joint best-fitting model. We do not observe any particular features in ACT; the SPT power is more suppressed at multipoles $\ell \gtrsim 1500$, but the points include an uncertain correlated extragalactic foreground contribution, whose dominant term is a Poisson shape. 

The addition of ACT and SPT helps WMAP constrain the basic six parameters due to a more precise determination of the higher order acoustic peak positions and amplitudes. The measurement of $\theta_A$ improves by a factor $2.2$ and the error on the baryon density is $1.6$ smaller compared to WMAP9 alone. However, as noted in \cite{Hou:2012xq}, the increased acoustic horizon scale leads to a predicted distance, $D_V$, to objects at redshift $z=0.57$, in units of the sound horizon at recombination, $r_s$, of $100r_s/D_V=7.66\pm0.14$, more than $2\sigma$ larger than measured by the BOSS experiment (\cite{Anderson:2012sa}, $100r_s/D_V=7.3\pm0.1$). The prediction at $z=0.35$, $100r_s/D_V=11.57\pm0.26$, is consistent at $1\sigma$ with the SDSS DR-7 observations (\cite{Padmanabhan:2012hf}, $100r_s/D_V=11.3\pm0.2$).
We find a preference for a scale-dependent primordial power spectrum  at $3.5 \sigma$ from the CMB, with $n_s = 0.9690 \pm 0.0089$ at 68\% confidence.

\begin{figure*}
\center
\includegraphics[width=\textwidth]{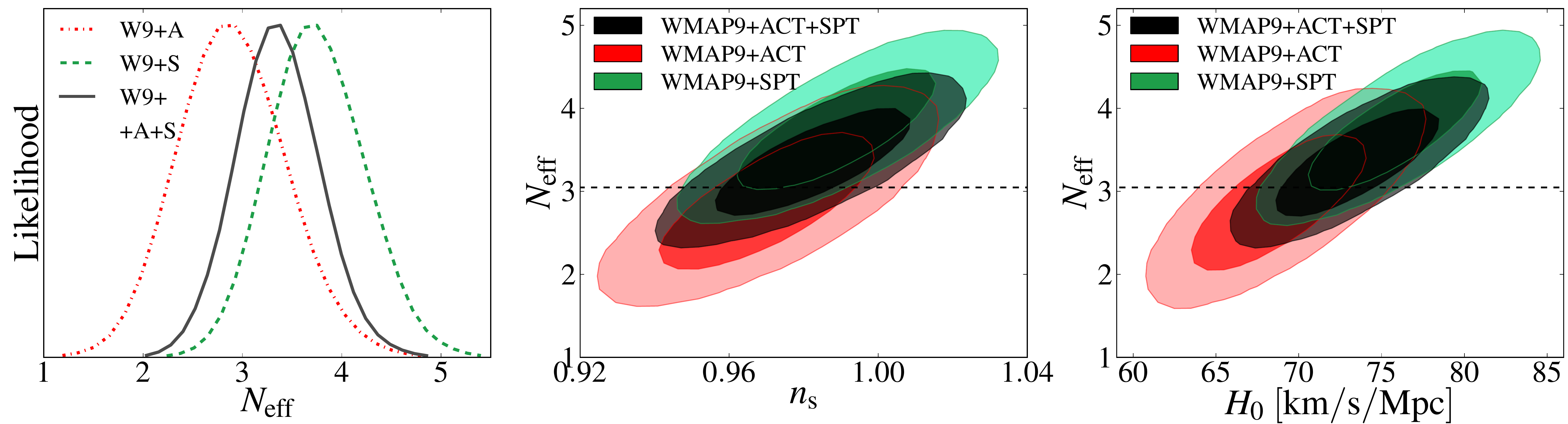}
\caption{\emph{Left:} Marginalized distribution of $N_{\rm{eff}}$ for different data combinations, showing consistency with three neutrino species. \emph{Middle and Right panels:} Marginalized $68\%$ and $95\%$ contours in the $N_{\rm{eff}}$- $n_s$ and $N_{\rm{eff}}$- $H_0$ planes; Neff is correlated with both parameters. The standard model expectation of $N_{\rm{eff}} = 3.046$ is indicated with dashed lines.
\label{fig:neff}
}
\end{figure*}
The CMB power spectrum is sensitive to the composition of the Universe. The radiation energy density is the energy density in photons plus the sum of the energy density in  relativistic species that do not couple electromagnetically, including standard model neutrinos.  We parametrize the energy density in other relativistic particles through $N_{\rm{eff}}$. In the standard cosmological model, $N_{\rm{eff}}=3.046$ \citep{dicus/etal:1982,lopez/etal:1998,mangano3046} describes the three known neutrino species. If there is an extra neutrino species that decouples at the same temperature as the standard neutrinos then $N_{\rm eff} \simeq 4$. If, instead, there is another light weakly interacting stable particle that decouples earlier, it will increase $N_{\rm eff}$ by the cube of the ratio of the decoupling temperatures.
The extra energy density in relativistic species has three noticeable effects on the CMB power spectrum \cite{Bashinsky:2003tk,Hou:2011ec,Archidiacono:2011gq,Smith:2011es}: (1) it increases the expansion rate of the Universe, which impacts both the acoustic and damping scale, an effect that is mostly degenerate with increasing the matter density, $\Omega_m h^2$; (2) it modulates the helium abundance from big bang nucleosynthesis, which in turn modifies the damping tail through free electrons available at recombination; and (3) the relativistic particles will free stream out of density fluctuations and suppress the amplitude of the power spectrum on small angular scales (an effect partially degenerate with increasing $n_s$, which enhances the amplitude of the power spectrum on those scales).

In our analysis, we vary $N_{\rm{eff}}$ as a free parameter and we assume that the same relativistic species are present at nucleosynthesis. We keep other quantities that describe the damping tail set to standard values: the total available electron abundance (determined by the primordial helium abundance $Y_p$) is consistent with standard big bang nucleosynthesis; the higher precision determination of recombination is used \cite{recomb11}, as implemented in \cite{camb}.
Combining the data we are considering in this work we find at $68\%$ confidence level: 
\begin{eqnarray}
N_{\rm{eff}} & =& 2.90 \pm 0.53 \quad \rm{(WMAP9+ACT)} \nonumber\\
N_{\rm{eff}} &=& 3.75 \pm 0.47 \quad \rm{(WMAP9+SPT)} \nonumber\\
N_{\rm{eff}} &=& 3.37 \pm 0.42 \quad \rm{(WMAP9+ACT+SPT)} .\nonumber
\end{eqnarray}
In Figure \ref{fig:neff} we show the distribution for $N_{\rm{eff}}$ from WMAP9 combined with ACT and SPT separately, and together. There is a $1.2 \sigma$ difference between the ACT and SPT estimates; as noted in \cite{Hou:2012xq}, the SPT data prefer a higher value, indicating more suppression of the small-scale spectrum. The probability that this variation is given by statistical scatter is around $50\%$, \cite{eigenfoot}. Based on the difference in the damping tail measurements, \cite{diValentino} decided to not combine the ACT and SPT data. In this paper we take a different approach and view the consistency sufficient for combination. As noted in \cite{feeney}, a Bayesian model comparison shows no evidence in favor of adding an additional $N_{\rm{eff}}$ parameter beyond those of the standard cosmology.

The correlations among $N_{\rm{eff}}$, $n_s$, and $H_0$ are also shown in Figure \ref{fig:neff}; the suppression of small scale power due to larger values of $N_{\rm eff}$  can be partially compensated by increasing $n_s$ and $\Omega_m h^2$. This leads to a larger derived value of $H_0$ if the CMB peak positions are held fixed.

Since a higher value for $N_{\rm eff}$ requires a higher matter density today, it gives a higher amplitude of gravitational potential fluctuations and an increased gravitational lensing signal. Measurements of the four-point function of the CMB temperature maps provide a measurement of the lensing deflection signal. The ACT \citep{Das:2013zf} and SPT \citep{vanEngelen/etal:2012} data constrain the amplitude of the lensing potential power spectrum at $\ell = 400$ to be $C_{400}^{\kappa \kappa}=(3.69 \pm 0.80)\times 10^{-8}$ for ACT and $C_{400}^{\kappa \kappa}=(2.92 \pm 0.54)\times 10^{-8}$ for SPT, yielding a combined $C_{400}^{\kappa \kappa} = (3.17 \pm 0.45)\times 10^{-8}$. 
Adding this constraint gives 
\begin{eqnarray}
N_{\rm{eff}} = 3.28 \pm 0.40 \quad \rm{(WMAP9+ACT+SPT+Lensing)} ,\nonumber
\end{eqnarray}
consistent with three neutrino species.


\section{Conclusions}\label{sec:conclusions}
Current microwave background power spectrum measurements are consistent
with the standard $\Lambda$CDM cosmological model, and independent data sets
are consistent with each other, with a mild tension between the ACT and SPT damping
tails. Upcoming maps from the Planck satellite will provide independent measurements
of the same sky regions with excellent foreground characterization.

\acknowledgements

\emph{Acknowledgements ---}This work was supported by the U.S. National Science Foundation through awards AST-0408698 and AST-0965625 for the ACT project, as well as awards PHY-0855887 and PHY-1214379. Funding was also provided by Princeton University, the University of Pennsylvania, and a Canada Foundation for Innovation (CFI) award to UBC. ACT operates in the Parque Astron\'omico Atacama in northern Chile under the auspices of the Comisi\'on Nacional de Investigaci\'on Cient\'ifica y Tecnol\'ogica de Chile (CONICYT). Computations were performed on the GPC supercomputer at the SciNet HPC Consortium. SciNet is funded by the CFI under the auspices of Compute Canada, the Government of Ontario, the Ontario Research Fund -- Research Excellence; and the University of Toronto. Funding from ERC grant 259505 supports EC, JD, and TL. We acknowledge the use of the Legacy Archive for Microwave Background Data Analysis (LAMBDA). Support for LAMBDA is provided by the NASA Office of Space Science.  
The likelihood code will be made public through LAMBDA (\url{http://lambda.gsfc.nasa.gov/}) and the ACT website (\url{http://www.physics.princeton.edu/act/}).



\begin{thebibliography}{50}
\bibitem{spergel94}
D.~N. Spergel, 1994 Warner Prize Lecture, Bull. Amer. Astron. Soc., 26, 1427 (1994).

\bibitem{jungman96}
  G.~Jungman, M.~Kamionkowski, A.~Kosowsky \& D.~N.~Spergel,  Phys.\ Rev.\ D, {\bf 54}, 1332 (1996).

\bibitem{zaldarriaga97}
  M.~Zaldarriaga, D.~N.~Spergel \& U.~Seljak, ApJ, {\bf 488},  1 (1997). 

\bibitem{bond97}
  J.~R.~Bond, G.~Efstathiou \& M.~Tegmark, MNRAS, {\bf 291}, L33 (1997).

\bibitem{Bennett:2012fp}
  C.~L.~Bennett, D.~Larson, J.~L.~Weiland, {et al.},  ArXiv e-prints,  {{\sffamily arXiv:1212.5225 [astro-ph.CO]}} (2012).
 
\bibitem{Hinshaw:2012fq}
  G.~Hinshaw, D.~Larson, E.~Komatsu, {et al.}, ArXiv e-prints,  {{\sffamily arXiv:1212.5226 [astro-ph.CO]}} (2012).

\bibitem{Das:2013zf}
  S.~Das, T.~Louis, M.~R.~Nolta, {et al.}, ArXiv e-prints,  {{\sffamily arXiv:1301.1037 [astro-ph.CO]}} (2013).

\bibitem{Dunkley:2013vu} 
J.~Dunkley, E.~Calabrese, J.~L.~Sievers, {et al.}, ArXiv e-prints,  {{\sffamily arXiv:1301.0776 [astro-ph.CO]}} (2013).

\bibitem{Sievers:2013wk}
  J.~L.~Sievers, R.~A.~Hlozek, M.~R.~Nolta, {et al.}, ArXiv e-prints,  {{\sffamily arXiv:1301.0824 [astro-ph.CO]}} (2013).

\bibitem{Story:2012wx}
  K.~T.~Story, C.~L.~Reichardt, Z.~Hou, {et al.}, ArXiv e-prints,  {{\sffamily arXiv:1210.7231 [astro-ph.CO]}} (2013).

\bibitem{Hou:2012xq}
  Z.~Hou, C.~L.~Reichardt, K.~T.~Story, {et al.}, ArXiv e-prints,  {{\sffamily arXiv:1212.6267 [astro-ph.CO]}} (2013).

\bibitem{Komatsu:2003fd}
  E.~Komatsu {et al.} [WMAP Collaboration],  ApJS, {\bf 148}, 119 (2003).

\bibitem{das/etal:2011}
 S. Das, B.~D. Sherwin, P. Aguirre, {et~al.}, PRL, {\bf 107}, 1301 (2011).

\bibitem{vanEngelen/etal:2012} A. van Engelen, R. Keisler, O. Zahn, {et al.}, ApJ, {\bf 756}, 142 (2012).

\bibitem{binning}
Here we use a slightly adapted binning of the spectra compared to those used in \cite{Dunkley:2013vu} and \cite{Sievers:2013wk}; this is preferable as it gives the two ACT regions more similar bandpower window functions. Using the new binning alters cosmological parameters by $\lesssim 0.1 \sigma$. The lensing amplitude reduces of about $0.2 \sigma$.

\bibitem{secondary}
In the secondary model there are 6 common parameters: two for the thermal and kinematic SZ effects, three for the CIB clustered and Poisson power, and a cross-correlation between tSZ and clustered CIB power. There is a separate Poisson radio power each for ACT and SPT; two Galactic parameters for ACT; and four calibration parameters for ACT, as described in \cite{Dunkley:2013vu}.

\bibitem{w7s12}
Using these SPT CMB-only bandpowers with the WMAP 7-year data we find the same cosmological parameters as in \cite{Story:2012wx}.
 This indicates that the foreground priors imposed in the SPT team's analysis (on the total SZ power, the clustered source power, and the Poisson source power), are consistent with the power that results from our multi-frequency analysis. It also further validates this new approach of pre-marginalizing over the foregrounds.

\bibitem{Lewis:2002ah}
A. Lewis \& S. Bridle, Phys.\ Rev.\ D, {\bf 66}, 103511 (2002). (Available from \texttt{http://cosmologist.info}).

\bibitem{Anderson:2012sa}
  L.~Anderson {et al.}, MNRAS,  {\bf 428}, 1036 (2013).
 
\bibitem{Padmanabhan:2012hf}
  N.~Padmanabhan, X.~Xu, D.~J.~Eisenstein, {et~al.}, {{\sffamily arXiv:1202.0090 [astro-ph.CO]}} (2012).
  
\bibitem{dicus/etal:1982}
D. Dicus, E.~W. Kolb, A.~M. Gleeson, {et~al.}, {\prd, {\bf 26}, 2694} (1982).

\bibitem{lopez/etal:1998}
R.~E. {Lopez}, S. {Dodelson}, A. {Heckler}, \& M.~S. {Turner}, {Phys.\ Rev.\ Lett., {\bf 82}, 3952} (1999).

\bibitem{mangano3046}
  G.~Mangano, G.~Miele, S.~Pastor, T.~Pinto, O.~Pisanti \& P.~D.~Serpico, Nucl.\ Phys.\ B, {\bf 729 } 221 (2005). 

\bibitem{Bashinsky:2003tk}
  S.~Bashinsky and U.~Seljak, Phys.\ Rev.\ D, {\bf 69}, 083002 (2004).

\bibitem{Hou:2011ec}
  Z.~Hou, R.~Keisler, L.~Knox, M.~Millea and C.~Reichardt, {{\sffamily arXiv:1104.2333 [astro-ph.CO]}} (2011).

\bibitem{Archidiacono:2011gq}
  M.~Archidiacono, E.~Calabrese and A.~Melchiorri, Phys.\ Rev.\ D {\bf 84}, 123008 (2011).

\bibitem{Smith:2011es}
  T.~L.~Smith, S.~Das and O.~Zahn, Phys.\ Rev.\ D {\bf 85}, 023001 (2012).
 
 \bibitem{recomb11} 
J. Chluba \& R. M. Thomas, MNRAS, 412, 748 (2011); Y. Ali-Haımoud \& C.~M.~Hirata, Phys. Rev. D, {\bf 83} 043513 (2011).

\bibitem{camb}
  A.~Lewis, A.~Challinor and A.~Lasenby, ApJ, {\bf 538}, 473 (2000).

\bibitem{eigenfoot} We perform an independent test of the degree of damping using the analysis of \cite{fbcs13}, testing deviations from standard recombination history \cite{recomb11}. \cite{fbcs13} shows that two ionization fraction eigenmodes can be measured using WMAP7+SPT and WMAP7+ACT data. The most sensitive mode mainly measures changes in the overall damping scale parameter, defining the shape of the $C_\ell$ damping envelope. We measure unity normalized mode amplitudes of $0.14 \pm 0.45$ for WMAP9+ACT, and $-0.80 \pm 0.37$ for WMAP9+SPT, indicating a mild tension for SPT. The combined WMAP9+ACT+SPT gives $-0.44 \pm 0.33$. The relative $C_\ell$ deviation from changing $Y_p$ looks similar to this dominant mode and has no tension with BBN, with $Y_p = 0.276\pm 0.022$ for WMAP9+ACT+SPT.

\bibitem{fbcs13} 
M. Farhang, J. R. Bond, J. Chluba \& E. Switzer, ApJ, 764, 137 (2013).

\bibitem{diValentino}
  E.~Di Valentino, S.~Galli, M.~Lattanzi, A.~Melchiorri, P.~Natoli, L.~Pagano and N.~Said,
 {{\sffamily arXiv:1301.7343 [astro-ph.CO]}} (2013).

\bibitem{feeney}
  S.~M.~Feeney, H.~V.~Peiris and L.~Verde,
  {{\sffamily  arXiv:1302.0014 [astro-ph.CO]}} (2013).

\end{thebibliography}
\end{document}